\documentstyle[prc,aps,epsfig]{revtex}
\newcommand{\be}{\begin{eqnarray}}
\newcommand{\ee}{\end{eqnarray}}

\newcommand{\bi}{\begin{itemize}}
\newcommand{\ei}{\end{itemize}}
\begin{document}
\twocolumn[\hsize\textwidth\columnwidth\hsize
           \csname @twocolumnfalse\endcsname
\title{Dynamical constraints on phase transitions}
\author{Klaus Morawetz$^{1,2}$}
\address{$^1$ LPC-ISMRA, Bld Marechal Juin, 14050 Caen, France\\
$^2$ GANIL, Bld Becquerel, 14076 Caen Cedex 5, France}
\maketitle
\begin{abstract}
The numerical solutions of nonlocal and local Boltzmann kinetic
equations
for the simulation of central heavy ion reactions are parameterized in terms
of time dependent thermodynamical variables in the Fermi liquid
sense. This allows one to discuss
dynamical trajectories in phase space. The nonequilibrium state is
characterized by non-isobaric, non-isochoric etc. conditions, 
shortly called iso-nothing conditions. Therefore a combination of
thermodynamical observables is constructed which allows one to locate
instabilities and points of possible phase transition in a dynamical
sense. We find two different mechanisms of instability, a short time
surface - dominated instability and later a spinodal - dominated volume
instability. The latter one occurs only if the incident energies do
not exceed significantly the Fermi energy and might be attributed to spinodal
decomposition. In contrast the fast surface explosion occurs far
outside the spinodal region and pertains also in the cases where the system
develops too fast to suffer a spinodal decomposition and where
the system approaches equilibrium outside the spinodal region. 
\end{abstract}
\vskip2pc]

\section{Introduction}

The collisions of heavy ions around the Fermi energy and the description
of multifragmentation phenomena fills an enormous literature. Mostly
the multifragmentation is attributed to a hypothetical liquid - gas phase
transition which is partially supported by mean field considerations
in equilibrium where the nonlinear density dependence of the
interaction energy leads to a liquid - gas like first order phase
transition. Therefore, the phenomena has been investigated in terms of
a spinodal decomposition. However this straightforward picture is
over-shadowed by at least two serious drawbacks. First we have to deal with
finite systems, where the phase transition appears modified and less
pronounced than in infinite bulk matter. And secondly, we have to face
the fact that the process evolves under extreme nonequilibrium
conditions. For a critical discussion of recent models on
multifragmentation see \cite{EH00}.
We want to investigate the latter two points here and will use a
microscopic approach which allows one to describe the time evolution of
the one-particle distribution function including binary
correlations. We will suggest a possibility of analyzing phase
transitions in terms of time dependent thermodynamical variables and
will be able, in this way, to see signals of instability in
nonequilibrium and finite systems. 

The standard treatment to investigate basic features of
multifragmentation processes is performed in terms of fluctuation
analysis starting from the Landau equation \cite{PR87,DDM90,DRS91}  
or BUU equations \cite{KH94,ACC95}. Observing that these kinetic
equations do not lead to enough fluctuations to describe
multifragmentation, additional stochasticity has been assumed and
incorporated resulting in Boltzmann-Langevin pictures
\cite{SASB90,AG90,RR90,ASBB92,ColBur93,CoCh94}. The large time
scale of fluctuations has been analyzed in \cite{Ch96}. It is found
that the large time evolution of the system is guided by cooperative
effects and fluctuations in a universal manner. The crucial role of
collision rate has been pointed out in that it enforces the diffusive regime.

We will adopt here a straight microscopic picture of kinetic theory
without additional fluctuations of Langevin sources. While this is
perfectly microscopical controlled we have to leave out the
possibility of describing fragment production. In contrast we will
investigate the thermodynamical trajectories as arising straight from
the solved kinetic equation as a Fermi
liquid. Therefore no coalescence or other cluster creation mechanisms
are used. This allows to restrict to the single particle distribution if
the two - particle correlations are included in the collision integral. This
is performed in the frame of the nonlocal kinetic theory. Since we want
to study the dynamical constraints of phase transitions as necessary
but not sufficient conditions we can expect already from the kinetic
theory an answer whether the system will undergo spinodal decomposition
or other forms of decomposition. In fact we will demonstrate that there
is a dominant surface emission at higher energies than the Fermi
energy while the spinodal decomposition can be accessed only for
energies lower or equal the Fermi energy. For higher energies the
system evolves too fast through the spinodal region to be influenced
sufficiently by spinodal decomposition. 

There are two experimental hints for two different regimes of
instability in heavy ion collisions around Fermi energy. The 
first one concerns the emulsion data recorded in the experiment by
F. Schussler et. al. which has been considered in
\cite{NINNP93}. There the fragments with charge $Z>2$ has been grouped
into two
different velocities, one around $0.16c$ and the other with $0.25c$. A
possible interpretation has been advocated that the higher velocity
group comes from fragments emitted at an early stage from the surface.  
The TDHF calculations seemed to support this picture. 

A second
experimental signal comes from the production of hard photons as
measured by the TAPS collaboration \cite{Sch97}. The extracted source
sizes by HBT interferometry has found to be too large if not two
sources are assumed. Moreover the calculated photon spectra shows a
clear prompt source of hard photons besides later thermal photons. The
latter second source vanishes for incident energies larger than
$60$MeV.
This indicates already that there is a transition between two
mechanisms of particle production and instability if the bombarding
energy exceeds 50-60 MeV.

We will show that indeed there can be identified two mechanisms; at
short times a
surface dominated emission and at later times a volume dominated spinodal decomposition. For low energies we will find that the volume
spinodal effects are visible while for higher energies only the
surface emission survives.

\section{Kinetic description}

We use for the description the recently derived
nonlocal kinetic
equation \cite{SLM96} for the one - particle distribution function
\begin{eqnarray}
\!\!\!&&{\partial f_1\over\partial 
t}+{\partial\varepsilon_1\over\partial {\bf k}}
{\partial f_1\over\partial 
{\bf r}}-{\partial\varepsilon_1\over\partial {\bf r}}
{\partial f_1\over\partial {\bf k}}
=\nonumber\\
&&\sum_b\int{d{\bf p}d{\bf q}\over(2\pi)^5\hbar^7}\delta\left(\varepsilon_1
+
\varepsilon_2-\varepsilon_3-\varepsilon_4+2\Delta_E\right)
\left|{\cal T}_{ab}\right|^2
=\nonumber\\
&&\Bigl[f_3f_4\bigl(1-f_1\bigr)\bigl(1-f_2\bigr)-
\bigl(1-f_3\bigr)\bigl(1-f_4\bigr)f_1f_2\Bigr],
\label{9}
\end{eqnarray}
with Enskog-type shifts of arguments \cite{SLM96}:
$f_1\equiv f_a({\bf k,r},t)$, $f_2\equiv 
f_b({\bf p,r}\!-\!\Delta_2,t)$,
$f_3\equiv 
f_a({\bf k\!-\!q}\!-\!\Delta_K,{\bf r}\!-\!\Delta_3,t\!-\!\Delta_t)$
, and
$f_4\equiv
f_b({\bf p\!+\!q}\!-\!\Delta_K,{\bf r}\!-\!\Delta_4,t\!-\!\Delta_t)$
. The
effective scattering measure, the ${\cal T}$-matrix is 
centered in all
shifts. The quasiparticle energy $\varepsilon$ contains 
the mean field
as well as the correlated self energy.

In agreement with \cite{NTL91,H90}, all gradient 
corrections are given
by derivatives of the scattering phase 
shift
\mbox{$\phi={\rm Im\ ln}{\cal 
T}^R_{ab}(\Omega,{\bf k,p,q,}t,{\bf r})$},
\begin{equation}
\begin{array}{lclrcl}\Delta_t&=&{\displaystyle
\left.{\partial\phi\over\partial\Omega}
\right|_{\varepsilon_1+\varepsilon_2}}&\ \ \Delta_2&=&
{\displaystyle\left({\partial\phi\over\partial {\bf p}}-
{\partial\phi\over\partial {\bf q}}-{\partial\phi\over\partial 
{\bf k}}
\right)_{\varepsilon_1+\varepsilon_2}}\\ &&&&&\\ 
\Delta_E&=&
{\displaystyle\left.-{1\over 
2}{\partial\phi\over\partial t}
\right|_{\varepsilon_1+\varepsilon_2}}&\Delta_3&=&
{\displaystyle\left.-{\partial\phi\over\partial {\bf k}}
\right|_{\varepsilon_1+\varepsilon_2}}\\ &&&&&\\ 
\Delta_K&=&
{\displaystyle\left.{1\over 2}{\partial\phi\over\partial 
{\bf r}}
\right|_{\varepsilon_1+\varepsilon_2}}&\Delta_4&=&
{\displaystyle-\left({\partial\phi\over\partial {\bf k}}+
{\partial\phi\over\partial 
{\bf q}}\right)_{\varepsilon_1+\varepsilon_2}}.
\end{array}
\label{8}
\end{equation}
After derivatives, 
$\Delta$'s are evaluated at the energy shell 
$\Omega\to\varepsilon_3+
\varepsilon_4$. Neglecting these shifts the usual BUU 
scenario appears.

The $\Delta$'s in the arguments of the distribution
functions in (\ref{9}) remind the non-instant and non-local
corrections in the scattering-in integral for classical 
particles.
The displacements of the asymptotic states are given
by $\Delta_{2,3,4}$. The time delay enters in an equal 
way the
asymptotic states 3 and 4. The momentum gain $\Delta_K$ 
also appears
only in states 3 and 4. Finally, there is the energy 
gain which is discussed 
in \cite{LSM99}. 
These nonlocal corrections to the usual Boltzmann 
equation are
a compact form of gradient corrections. It ensures that 
the
conservation laws contain besides the mean-field 
correlations also the
two particle correlations.

Despite its complicated form it is possible to solve 
this kinetic equation 
with standard Boltzmann numerical codes and to implement 
the shifts \cite{MLSCN98}.
 Therefore we have calculated the shifts for different 
realistic nuclear potentials
 \cite{MLSK98}. The numerical solution of the nonlocal 
kinetic equation 
has shown an observable effect in the dynamical particle 
spectra of around $10\%$. 
The high energetic tails of the spectrum are enhanced due 
to more energetic two-particle collisions 
in the early phase of nuclear collision. Therefore the 
nonlocal corrections lead 
to an enhanced production of preequilibrium high 
energetic particles.

Besides the nonlocal shifts and cross section which has been
calculated from realistic potentials we adopt here the view that the
selfenergy $\varepsilon$ is parameterized in terms of the Skyrme potential for
which we
use a soft potential of the form
\be
\varepsilon={p^2\over 2 m} + A \left ({n\over n_0}\right )+B \left
  ({n\over n_0}\right )^\sigma.
\label{hf}
\ee
For a derivation of collision integrals and the Skyrme potential
(\ref{hf}) from the same microscopic footing, see \cite{M00}.

\subsection{Balance equations}
By multiplying the kinetic equation with $1,{\bf p},\varepsilon$ one obtains the
balance for the particle density $n$, the momentum density $J$ and the
energy density ${\cal E}$. 
Without nonlocal corrections the collision
integrals vanish for the density and momentum balance and we get the standard
balance equations for the quasiparticle parts
\be
&&{\partial n_a({\bf r},t)\over \partial t}+{\bf {\partial \over \partial r}} \int
{d {\bf p}\over (2 \pi \hbar)^3} {\partial \varepsilon\over \partial {\bf p}}  f_a({\bf p,r},t)=0\nonumber\\
&&{\partial J_i({\bf r},t)\over \partial t} +{\partial \over \partial
  r_j} {\cal P}^{\rm qp}_{ij}=0
\ee
with the quasiparticle density, the current and the momentum tensor
\begin{eqnarray}
n_a^{}&=&\int{d{\bf p}\over(2\pi\hbar)^3}f_a
\nonumber\\
{\bf J}({\bf r},t)&=&\int {d {\bf p}\over (2\pi\hbar)^3} \, {\bf p}\, f_a({\bf p,r},t)\nonumber\\
{\cal P}_{ij}^{\rm qp}&=&\sum_a\int{d{\bf p}\over(2\pi\hbar)^3}\left(p_j
{\partial\varepsilon_a\over\partial p_i}+
\delta_{ij}\varepsilon_a\right)f_a-
\delta_{ij}{\cal E}^{\rm qp}
\label{10a}
\end{eqnarray}
where the quasiparticle energy is given by
\be
{\cal E}^{\rm qp}
&=&\sum_a\int{d{\bf p}\over(2\pi\hbar)^3}{p^2\over 2m}f_a
\nonumber\\
&+&{1\over 2}\sum_{a,b}\int{d{\bf k}d{\bf p} \over(2\pi\hbar )^6}
{\cal T}_{ab}(\varepsilon_1+\varepsilon_2,{\bf k,p},0)f_af_b
\label{eqp}
\ee
and the pressure is as usual
\be
P=\frac 1 3 {\cal P}_{ii}.
\ee
The quasi particle energy of the system varies as 
\be
\delta {\cal E^{\rm qp}}&=&\int
{d {\bf p}\over (2 \pi \hbar)^3} {\delta {\cal E}^{\rm qp}\over \delta
    f({\bf p,r},t)} \delta f({\bf p,r},t)
\nonumber\\
&=&\int
{dp\over (2 \pi \hbar)^3} \varepsilon \delta f({\bf p,r},t)
\ee
and since we adopt the parameterization of quasiparticle energy
(\ref{hf}),
the quasiparticle part of the total energy density reads
\be
&&{\cal E}^{\rm qp}({\bf r},t)=\sum\limits_a\int{d {\bf p}\over (2\pi \hbar)^3}
{p^2\over 2 m} f_a({\bf p,r},t)\nonumber\\
&&+{A} {\textstyle \frac{n^2({\bf r},t)}{2 n_0}} 
      + B{\textstyle \frac{n({\bf
            r},t)^{\sigma+1}}{(\sigma+1)n_0^\sigma}}
+{\cal E}^{\rm Born}.
\label{en}
\ee
Please note that besides the mean field (\ref{hf}) we have also a Born
correlation term ${\cal E}^{\rm Born}$ coming from the second term of
(\ref{eqp}), see \cite{MK97},
\be
{\cal E}^{\rm Born}(t)&=&{ {\cal E}_F^2} {2
  \log 2-11\over 70 \pi^3} {m\over \hbar^2}{\sigma} +o(T^3).
\ee
The balance of the quasiparticle part of the energy density reads
from the kinetic equation
\be
&&{\partial {\cal E}^{\rm qp}({\bf r},t) \over \partial t} +{\bf {\partial \over \partial r}} \sum\limits_a\int
{d{\bf p}\over (2 \pi \hbar)^3} \varepsilon_a {\partial \varepsilon_a\over \partial {\bf p}}
f_a({\bf p,r},t)=0.
\label{e1}
\ee

The correlational parts of the density, pressure and energy are
coming from genuine two-particle correlations beyond Born
approximation which are also derived from the balance equations of
nonlocal kinetic equations \cite{SLM96}. It has been shown that they
establish the complete conservation laws. 
These $\Delta$-contributions following from the nonlocality of the
scattering integral read for the energy, pressure tensor and density
\begin{eqnarray}
{\cal E}_c&=&{1\over 2}\sum_{a,b}\int{d{\bf k}d{\bf p}d{\bf
    q}\over(2\pi\hbar)^9\hbar} \, \Psi \,
[\varepsilon_1+\varepsilon_2]\Delta_t,
\nonumber\\
{\cal P}^c_{ij}&=&\!{1\over 2}\!
\sum_{a,b}\!\!\int\!{d{\bf k}d{\bf p}d{\bf q}\over(2\pi\hbar)^9\hbar} \Psi
\left[(p\!+\!q)_i\Delta_{4j}\!+\!(k\!-\!q)_i\Delta_{3j}\!-\!p_i\Delta_{2j}\right],
\nonumber\\
n_c&=&\sum_b\int{d{\bf k}d{\bf p}d{\bf q}\over(2\pi\hbar)^9\hbar} \Psi \Delta_t,
\label{10}
\end{eqnarray}
where $\Psi=|{\cal T}_{\rm ab}^R|^22\pi\delta(\varepsilon_1\!+\!\varepsilon_2\!-
\!\varepsilon_3\!-\!\varepsilon_4)f_1f_2(1\!-\!f_3\!-\!f_4)$ is the
probability to form a molecule during the delay time $\Delta_t$.

While these correlated parts are present in the numerical results and
can be shown to contribute to the conservation laws we will only discuss the
thermodynamical properties in terms of quasiparticle quantities to
compare as close as possible with the mean field or local BUU
expressions. The discussion of these correlated two - particle
quantities are devoted to a separate consideration.

\subsection{Dynamical thermodynamical variables}
We want now to construct the time dependent global termodynamical
variables.
From the distribution function $f({\bf p,r},t)$ the local density, current
and energy densities
are given by
\be
{n}({\bf r},t)&=&\displaystyle{\int{d {\bf p}\over (2\pi \hbar)^3}{f({\bf p,r},t)}}\nonumber\\
{\bf J}({\bf r},t)&=&\displaystyle{\int{d {\bf p}\over (2\pi \hbar)^3} {\bf p} { f({\bf p,r},t)}}\nonumber\\
{\cal E}_K({\bf r},t)&=&\displaystyle{\int{d {\bf p}\over (2\pi \hbar)^3} {p^2\over 2 m} {f({\bf p,r},t)}}
\label{var}
\ee
which are computed directly from the numerical solution of the kinetic
equation in terms of test particles. Please note that the above
kinetic energy includes the Fermi motion.

\subsubsection{Temperature}

The global variables per particle number like kinetic energy, Fermi
energy and collective energy are obtained by spatial integration
\be
{\cal E}_K(t)&=&{\displaystyle{\int d{\bf r} \,{\cal E}_K({\bf
      r},t)}\over \displaystyle{\int d{\bf r} \, {n}({\bf r},t)}}
\nonumber\\
{\cal E}_F(t)&=&{\displaystyle{\int d{\bf r} \, {\cal E}_f[{n}({\bf
    r},t)]{n}({\bf
  r},t)} \over \displaystyle{\int d{\bf r} \, {n}({\bf r},t)}}
\nonumber\\
{\cal E}_{\rm coll}(t)&=&{\displaystyle{\int d{\bf r} \,{{J}({\bf r},t)^2
    \over é m \,{n}({\bf
  r},t)}} \over \displaystyle{\int d{\bf r} \, {n}({\bf r},t)}}
\ee
where we have used the local density approximation \cite{SHJGRSS89}.
Now we adopt the picture of Fermi liquid theory which connects the
temperature with the kinetic energy as
\be
{\cal E}_K(t)=\frac 3 5 {\cal E}_F(t)+ {\cal E}_{\rm coll}(t) +
{\pi^2\over 4 {\cal E}_F(t)} {
  T(t)}^2
\label{t}
\ee
from which we deduce the global temperature. The definition of
temperature is by no means obvious since it is in principle an
equilibrium quantity. One has several possibilities
to define a time dependent equivalent
temperature which
should approach the equilibrium value when the system approaches
equilibrium. In \cite{GW00,GWF00} the
definition of slope temperatures has been discussed and compared to local space
dependent temperature fits of the distribution function of
matter. This seems to be a good measure for higher energetic
collisions in the relativistic regime. 
Since we restrict here to collisions in the Fermi energy domain and do not want to add coalescence models we will not use
the slope temperature. Moreover we define the global temperature in
terms of global energies which are obtained by local quantities rather
than
defining a local temperature itself. This has the advantage that we do
not consider local energy fluctuations but only a mean evolution of temperature. 

\subsubsection{Energy and pressure}

The mean field part of the energy is given by
\be
U(t)&=&{\cal E}^{\rm qp}(t)-{\cal E}_K(t)-{\cal E}^{\rm
  Born}(t)\nonumber\\
&=&{\displaystyle{\int d{\bf r} \left (A{{ n({\bf r},t)}^2 \over 2 n_0}
    +B{ {n({\bf r},t)}^{s+1} \over (s+1) n_0^s}\right )}\over
\displaystyle{\int d{\bf r} \, {n}({\bf r},t)}}
\ee
from which one deduces the pressure per particle
\be
P(t)&=&\frac 2 3 ({\cal E}_K(t)-{\cal E}_{\rm coll}(t))+ \frac 4 3 {\cal E}^{\rm
  Born}(t)\nonumber\\
&+&{\displaystyle{\int d{\bf r} \left (A{ { n({\bf r},t)} \over 2 n_0}
    +B{ s \,{ n({\bf r},t)}^{s} \over (s+1) n_0^s}\right )} \over \displaystyle{\int d{\bf r}\, { n}({\bf r},t) }}.
\ee

In order to compare now the local BUU with the nonlocal BUU scenario we
consider the energy which would be the total energy in the local BUU
without Coulomb energy
\be
{\cal E}(t)={\cal E}_K(t)-{\cal E}_{\rm coll}(t)+U(t).
\label{e}
\ee

This expression does not contain the two - particle correlation energy
which is zero for BUU and the Coulomb energy. The reason for
considering this energy for dynamical trajectories is that we want to
follow the trajectories in the picture of mean field and usual spinodal
plots.

\subsubsection{Density}

\begin{figure}
\parbox[]{17cm}{
\parbox[]{8cm}{
\psfig{file=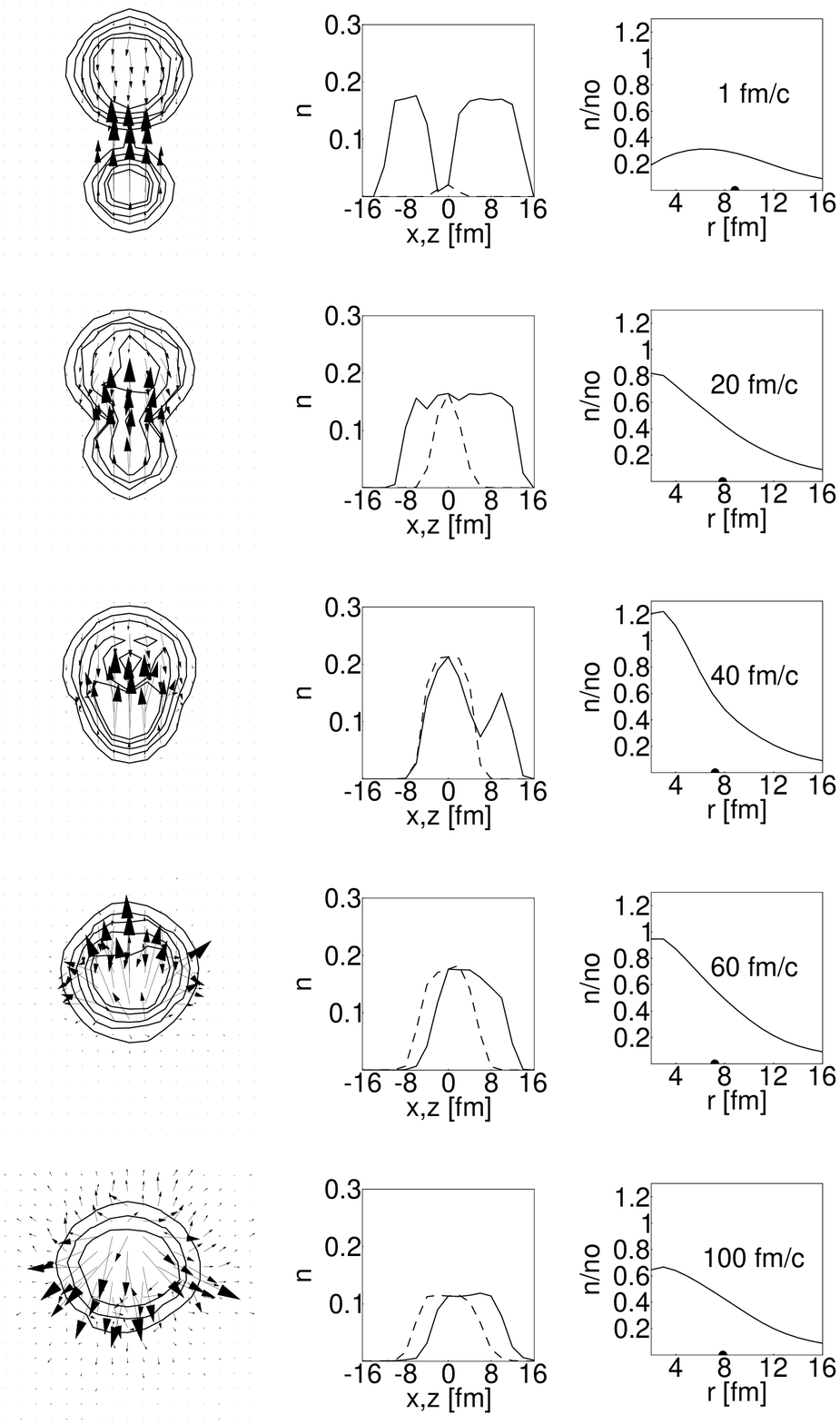,height=10cm,angle=0}
}
\parbox[]{8cm}{
\psfig{file=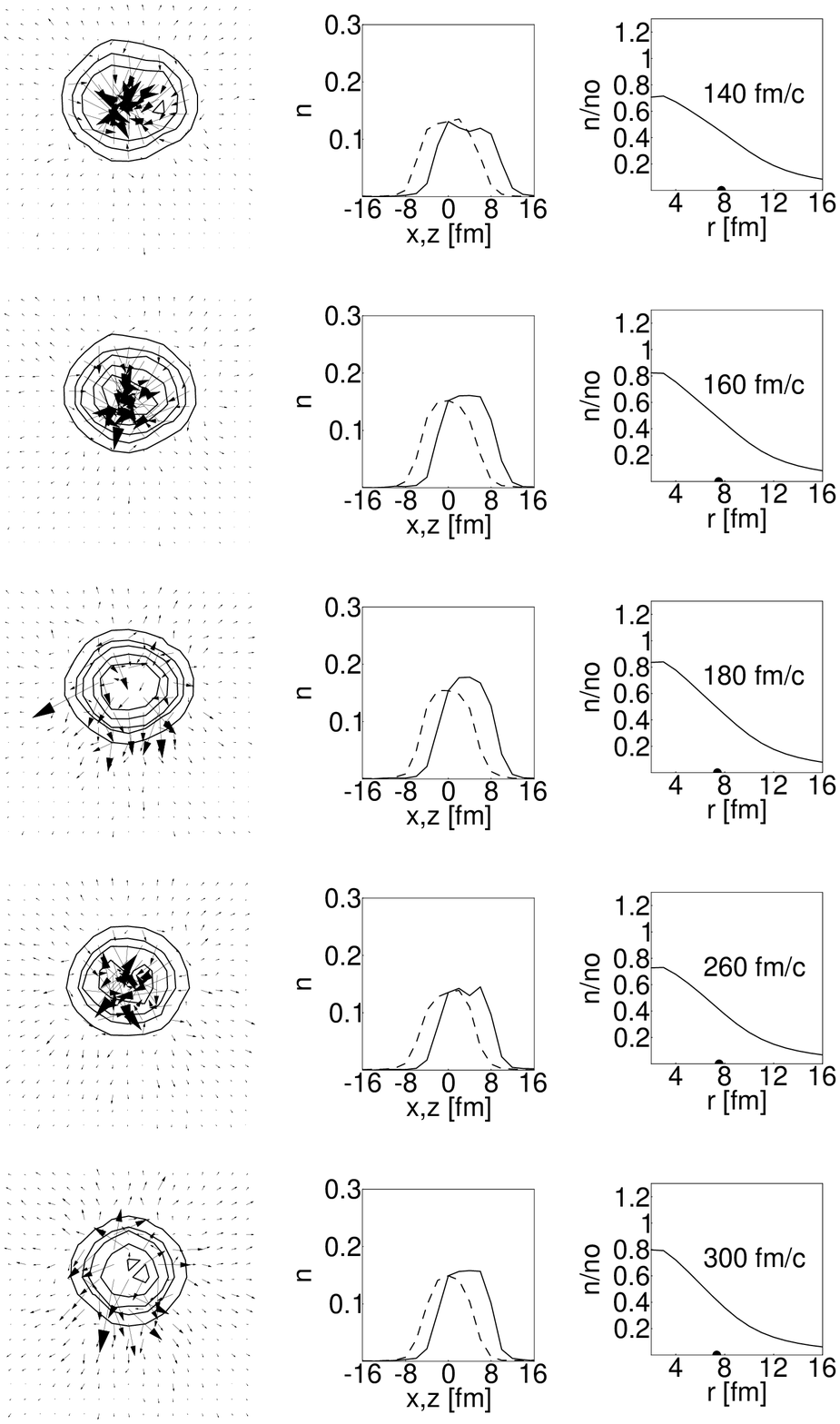,height=10cm,angle=0}
}
}
\vspace{2ex}
\caption{The time evolution of central collisions 
$Ni+Au$ at
  $25$MeV per nucleon. The density contours in the spatial plane $(x,0,Z)$ is
  plotted in the left figure in the range $-15,15$fm and the arrows characterize the values of
  the local current $J$ according to (\protect\ref{var}). The middle
  figure gives the density profile in ${\rm fm}^{-3}$ of the beam direction (solid line) 
and perpendicular to the beam (dotted line). The corresponding right panel shows the global
  density ratio to nuclear density $no=0.16{\rm fm}^{-3}$ defined in a sphere versus the radius of the sphere. The
  mean square radius is marked explicitly by a dot on the radius axes.}
\label{niau25n}
\end{figure}

\begin{figure}
\parbox[]{17cm}{
\parbox[]{8cm}{
\psfig{file=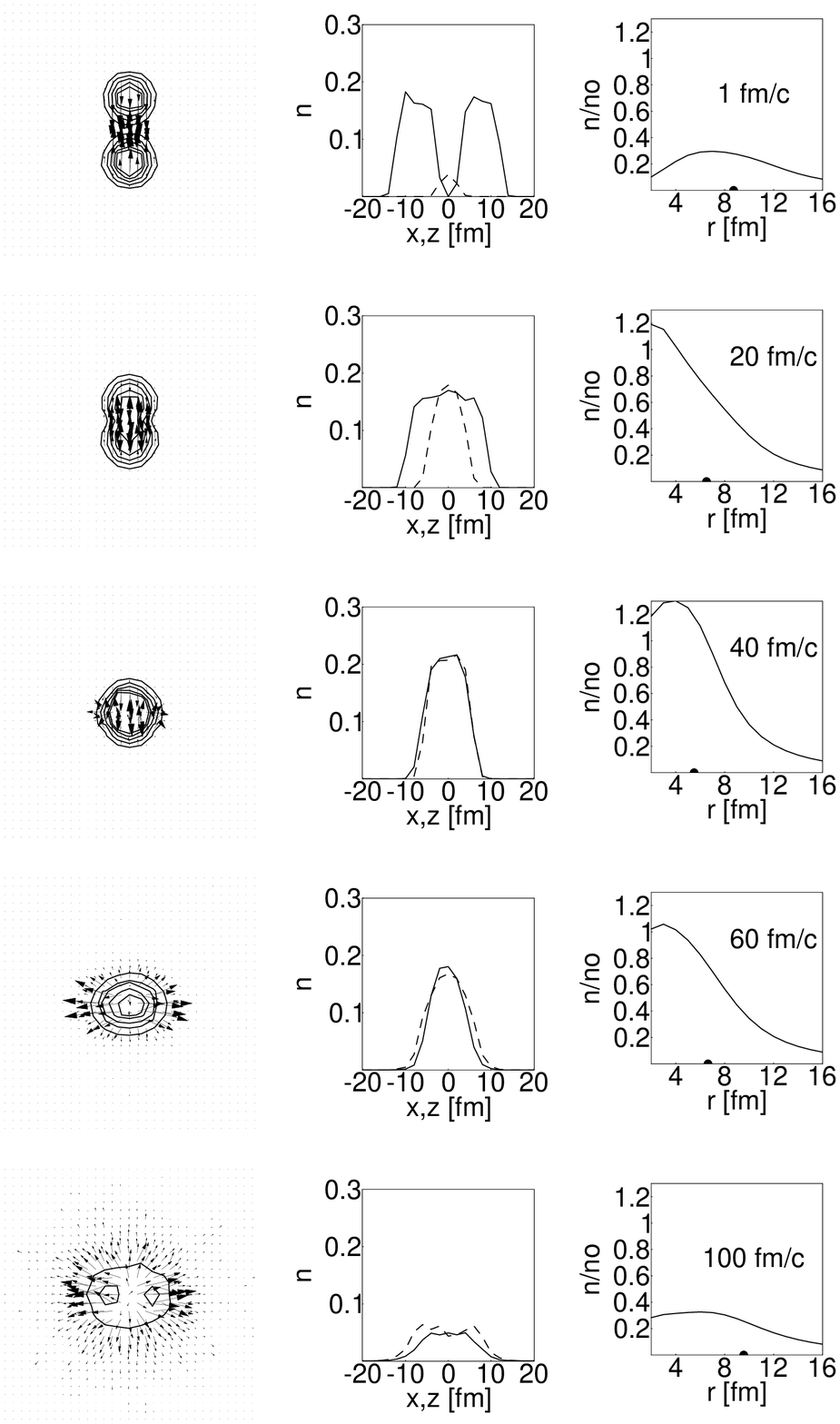,height=10cm,angle=0}
}
\parbox[]{8cm}{
\psfig{file=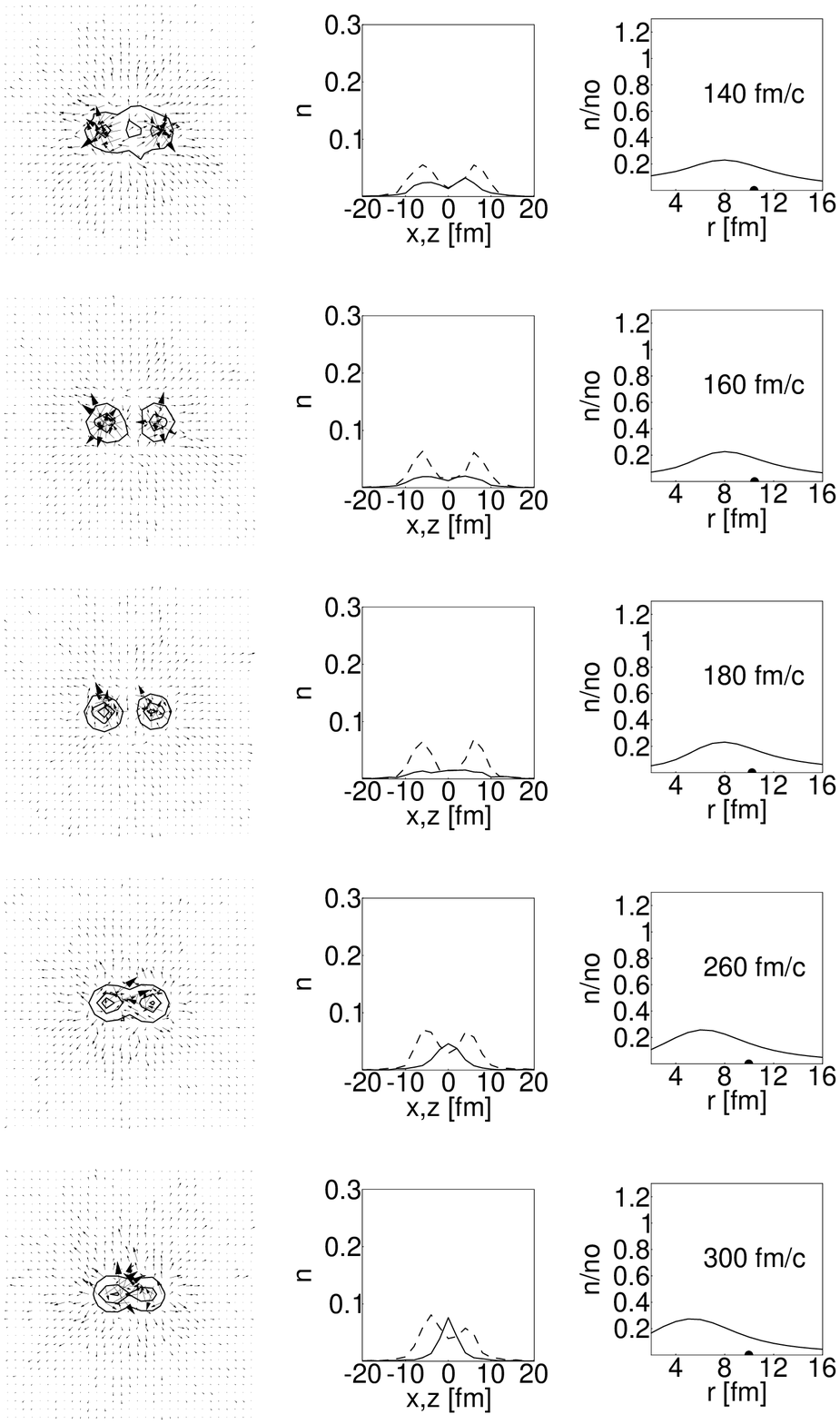,height=10cm,angle=0}
}
}
\vspace{2ex}
\caption{The time evolution of central collisions  
$Xe+Sn$ at
  $50$MeV per nucleon analogous to figure \protect\ref{niau25n}.}
\label{xesn50n}
\end{figure}

To define the density exhibits to some extend a problem. To illustrate
this fact we have plotted in figure \ref{niau25n} and \ref{xesn50n}
the density evolution. We see that depending on the considered volume sphere
we obtain different global densities. We follow here the point of view
that the mean square radius will be used as a sphere 
to define the global density. This is also supported by the
observation that the mean square radius follows the visible
compression. This becomes evident in figure \ref{xesn50n} for a
symmetric reaction at higher energies where at $40$fm/c we see a clear
compression. If we define the volume by a density cut $n>n_0/10$
in spatial domain we will not see
compression at all since the matter is
evaporating and this volume increases correspondingly to compression.  
Therefore we think that the sphere with the mean square radius is a good compromise.

\section{Iso-nothing conditions in equilibrium}

Let us first recall the figures of mean field isotherms in
equilibrium. 
The mean field Skyrme and Born correlational energy is 
\be
{\cal E}=\frac 3 2 n T {f_{5/2}\over f_{3/2}}+{A\over 2 n_0} n^2+{B
  \over (s+1)n_0 } n^{s+1}+{\cal E}^{\rm Born}
\label{hf1}
\ee
with the kinetic energy in terms of
standard Fermi integrals and the density
\be
n={g\over \lambda^3} f_{3/2}
\ee
with $g$ the spin, isospin,... degeneracy.
The corresponding pressure reads
\be
P&=&n^2 {d ({\cal E}/n) \over d n} \nonumber\\
&=&n T {f_{5/2} \over f_{3/2}}+{A\over 2 n_0} n^2+{B s\over (s+1)n_0 } n^{s+1}.
\ee

\begin{figure}
\noindent\psfig{file=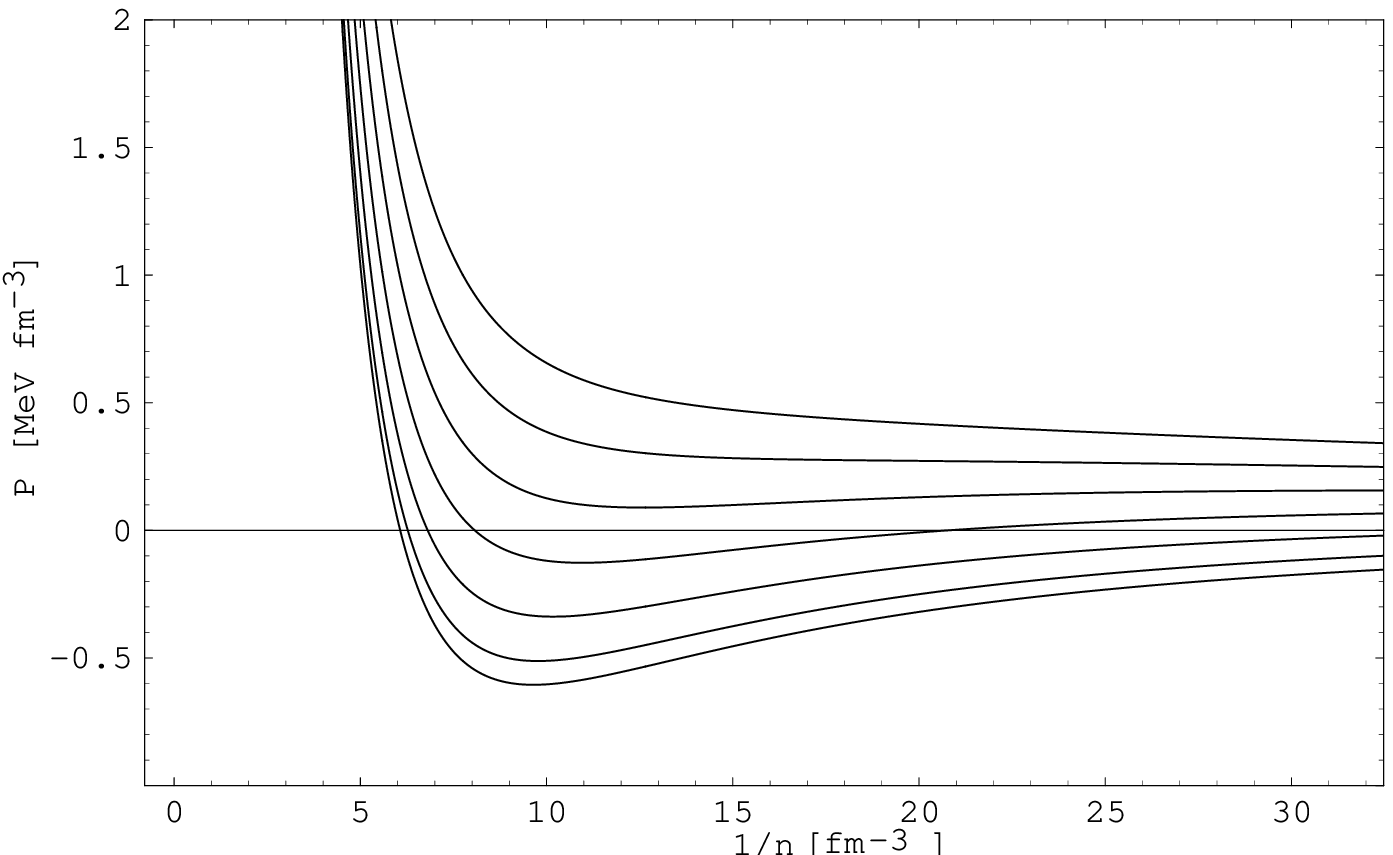,width=8cm}
\noindent\psfig{file=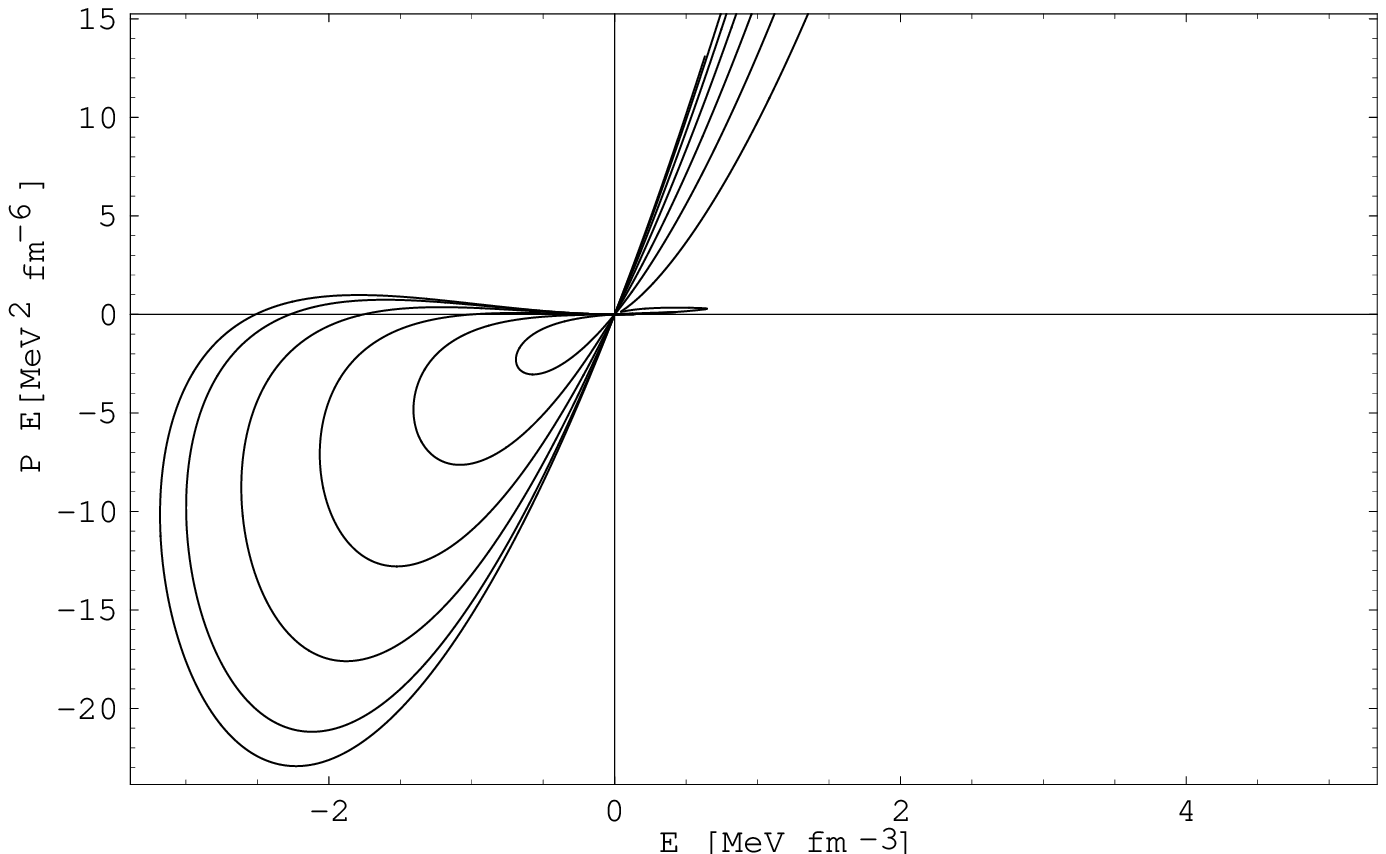,width=8cm}
\vspace{2ex}
\caption{The isotherms for the pressure density versus volume (above) and for
  the product of pressure and energy density versus energy density
  (below). The temperatures are $T=1,4,7,10,13,16,19$MeV.}\label{1}
\end{figure}

We obtain the typical van der Waals curves in figure
\ref{1}. Since we have neither isothermal nor isochoric nor isobaric
conditions in simulations, shortly since we have iso-nothing conditions, we
have to find a representation of the phase transition curves which are
independent of temperature but which reflects the main features of phase
transitions. This can be achieved by the product of energy and
pressure density versus energy density in figure \ref{1} below. This
plot shows that all instable isotherms exhibit a minimum in the left
lower quarter. There the energy is negative denoting bound state
conditions but
the pressure is already positive which means the system is
unstable. The first isotherm above the critical one does not touch
this quarter but remains in the right upper quarter where the energy
and pressure are both positive and the system is expanding and
decomposing unboundly. The left upper quarter denotes negative pressure
and energy indicating that the system is bound and stable.

\onecolumn

\begin{figure}
\psfig{file=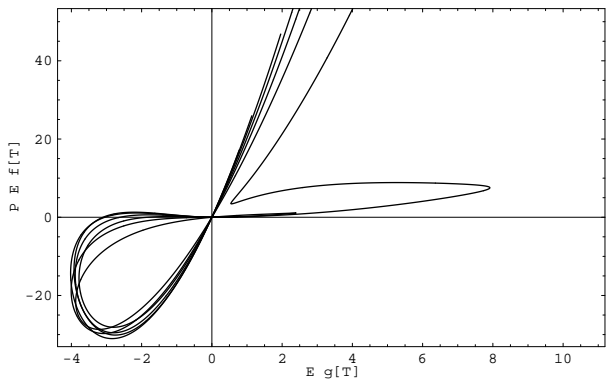,width=8cm}
\vspace{2ex}
\caption{The isotherms of the product of pressure and energy density
  versus energy density scaled by a temperature polynomial $f(T)$, $g(T)$. 
The temperatures are $T=1,4,7,10,13,16,19$MeV respectively. All critical isotherms
collapse on one line in the left lower quarter.}\label{2}
\end{figure}

\twocolumn
In order to achieve now a temperature independent plot we scale both
axes of figure \ref{1} (below) with a temperature dependent polynomial
and achieve that all critical isotherms are collapsing on one curve in
the left lower quarter, see figure \ref{2}. 
The first isotherms above
the critical one does not enter the left lower quarter. We consider
this scaling as adequate for iso-nothing conditions. A phase
transition should be possible to observe if there occurs a minimum in
the left half of this plot at negative energies. 

The idea of plotting combinations of pressure and energy is
similar to the one of softest point \cite{HS95} in analyzing QCD phase
transitions. There the simple pressure over energy ratio leads to a
temperature independent plot due to ultrarelativistic energy -
temperature relations. In our case we have a Fermi liquid behavior at
low temperatures and have to scale differently. In particular we have used
in figure \ref{2} the temperature dependent polynomials 
\be
g[x=T/{\rm MeV}]&=&1.2+\frac 1 3 10^{-1} x+\frac 1 3 10^{-3} x^2+\frac
1 5 10^{-4}x^4\nonumber\\
&&+ \frac
1 8 10^{-6} x^6
\nonumber\\
f[x=T/{\rm MeV}]&=&1.2+\frac 1 3 10^{-2} x+\frac 1 3 10^{-2}
x^2+10^{-5}x^4\nonumber\\
&&+ 10^{-7} x^6
\ee
which are producing a temperature independent plot in figure \ref{2} for the specific
used mean field potential parameterization.

\section{Nonequilibrium thermodynamics}

\begin{figure}
\centerline{\psfig{file=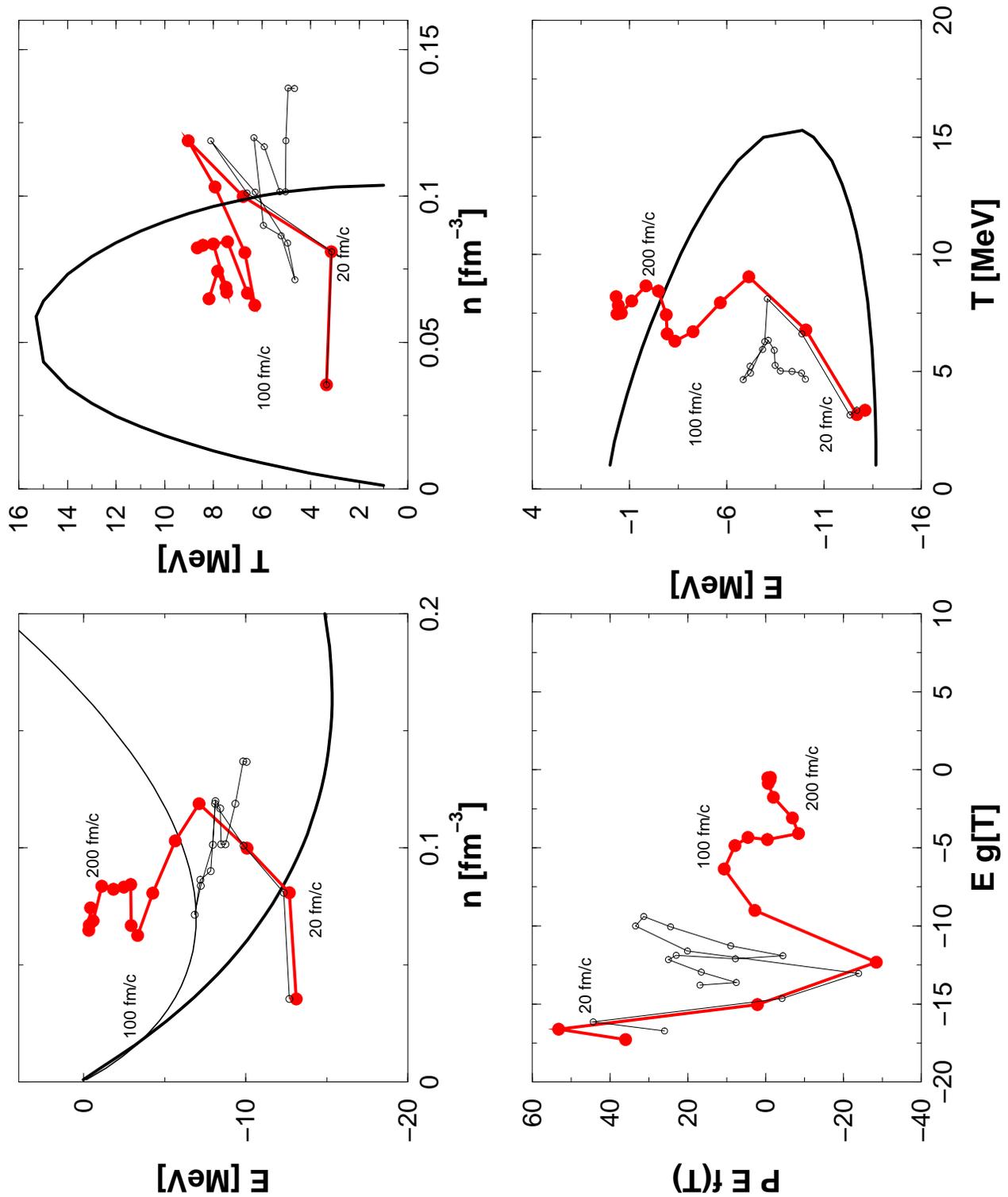,height=20cm,angle=0}}
\caption{The dynamical trajectories of the energy (\protect\ref{e}),
  density and
  temperature (\protect\ref{t}) in the nonlocal (gray thick) and in
  the local
  BUU (black thin) scenario. The considered reaction is $^{129}Xe$ on $^{119}Sn$ at $25$MeV
  lab energy. The dots mark the times in steps of $20$fm/c up to
  total of $300$fm/c. To guide the eye the zero temperature mean field
  energy (thick line) and the pressure (thin line) is plotted in the upper
  left picture and in the right figures the
  spinodal line for infinite matter is given. The scaled
  combinational plot analogous to figure \protect\ref{2} is given in
  the left lower plane.}\label{xesn25}
\end{figure}

Let us now inspect the dynamical trajectories for the above defined
temperature, density and energy. In figure \ref{xesn25} we have
plotted the dynamical trajectories for a charge - symmetric reaction
of $Xe$ on $Sn$ at $25$MeV lab - energy. The solution of the nonlocal
kinetic equation is
compared to the local BUU one. One sees in the temperature versus density
plane that the point of highest compression is reached around
$60$fm/c with a temperature of $9$ MeV. 

After this point of highest
overlap or fusion phase we have an
expansion phase where the density and temperature is decreasing. While
the compression phase is developing similar for the BUU and for the nonlocal
kinetic equation we see now differences in the development. First the
temperature of the nonlocal kinetic equation is around $2$MeV higher than
the local BUU result. This is due to the release of correlation
energy into kinetic energy which is not present in the local BUU
scenario. After this expansion stage until times of $120$fm/c we see that the
BUU trajectories come to a rest inside the spinodal region while the
nonlocal scenario leads to a further decay. This can be seen by
the continuous decrease of density and increase of
energy. Since matter is more decomposed with the nonlocal kinetic equation
we also heat the system more due to Coulomb acceleration. This
leads to the enhancement of temperature compared to BUU. An
oscillating behavior occurs at later
times which reflects an interplay between short range correlation and long
range Coulomb repulsion. The decomposition leads almost to free
gaseous matter after $300$fm/c as can be seen in the energy versus
density plot. 

Please note that although the trajectories seems to
equilibrate inside the spinodal region when one considers the temperature
versus density plane, we see that in the corresponding energy versus
temperature plane the trajectories travel already outside the
spinodal region. This underlines the importance to investigate the region of
spinodal decomposition in terms of a three dimensional plot instead
of a two dimensional one like in the recently discussed caloric curve
plots. Different experimental situations lead to different curves as
long as the third coordinate (pressure or density) remains
undetermined.

The iso-nothing plot analog to figure \ref{2} in the lower left corner shows that the point
of highest compression is linked to a first instability seen as a
pronounced minimum of the trajectory in the left quarter. This is
connected with a pronounced surface emission and connected with
anomalous velocity profiles
\cite{MTP00}. We will call this phase {\it surface emission} instability further on.
At $180$ fm/c we see a second minimum which is taking place inside
the spinodal region. This instability we might now attribute to spinodal
decomposition since the trajectories developing slower and remain
inside the spinodal region. The BUU shows the same qualitative minima
but the
matter rebounds and the trajectories move towards negative energies
again. In opposition the nonlocal scenario leads to a further
decomposition of matter as described above.

In the next figure \ref{xesn33} we have plotted the same reaction as
in figure \ref{xesn25} but at higher energy of $33$MeV. We recognize a
higher compression density and temperature than compared to the lower
bombarding energy. Consequently the trajectories develop further towards
the unbound region of positive energy after $300$fm/c. While the first
surface emission instability is strongly pronounced we see that the
second minimum in the iso-nothing plot is already weaker indicating
that the role of spinodal decomposition is diminished. The
trajectories in the temperature versus density plot comes still in the
spinodal region at rest but travels already outside the spinodal
region if the
energy versus temperature plot is considered. This shows that the
trajectories start to develop too fast to suffer
much spinodal decomposition. 

\begin{figure}
\centerline{\psfig{file=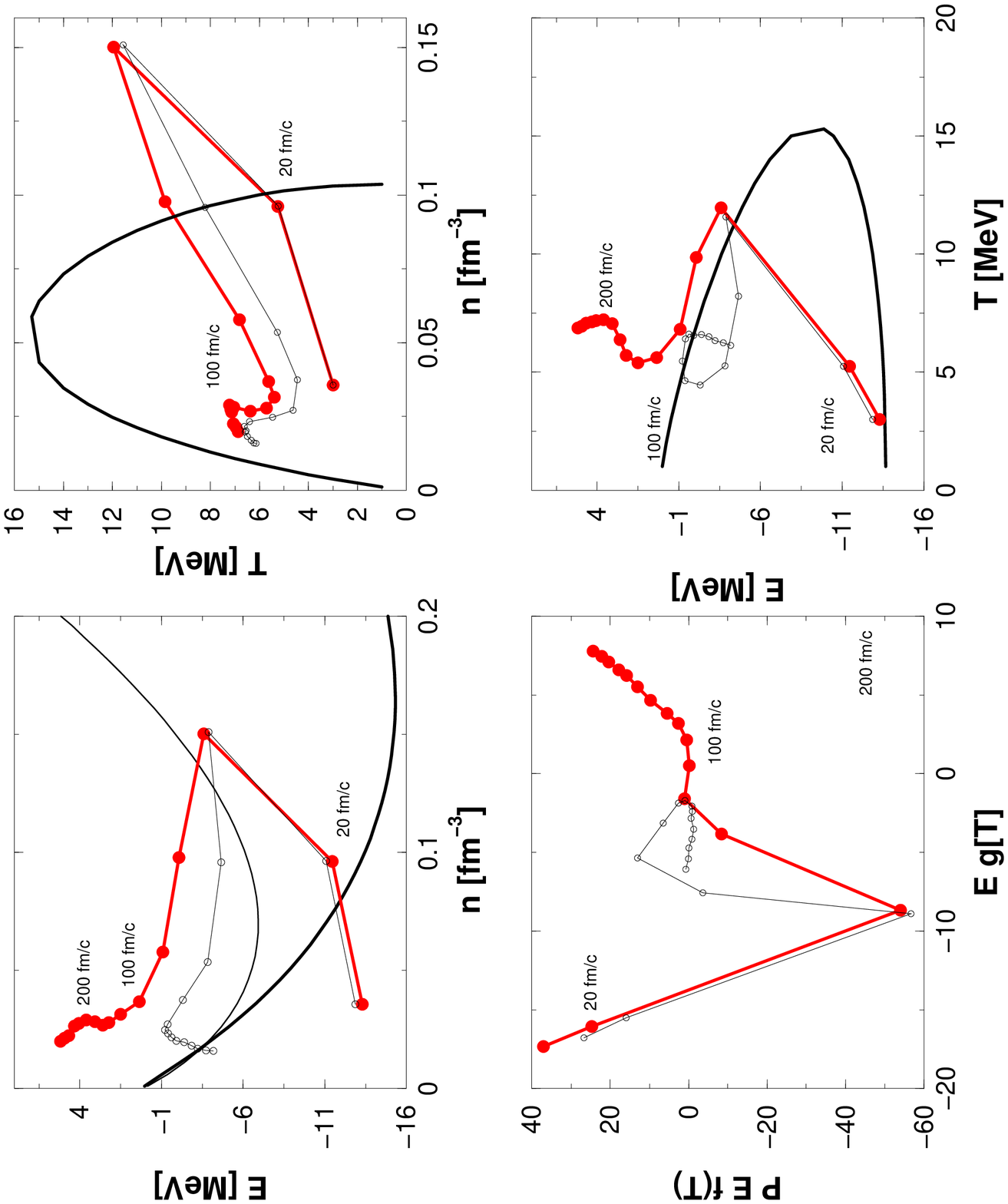,height=20cm,angle=0}}
\caption{The same figure as \protect\ref{xesn25} but for $50$MeV lab energy.}\label{xesn50}
\end{figure}

If we now plot the same reaction at $50$MeV in figure \ref{xesn50} we
see that the trajectories come at rest outside the spinodal region whatever
plot is used
and no second minima is seen anymore in the iso-nothing plot. But, the
surface emission instability is still very pronounced and is probably here the
leading mechanism of matter disintegration.  

\begin{figure}
\centerline{\psfig{file=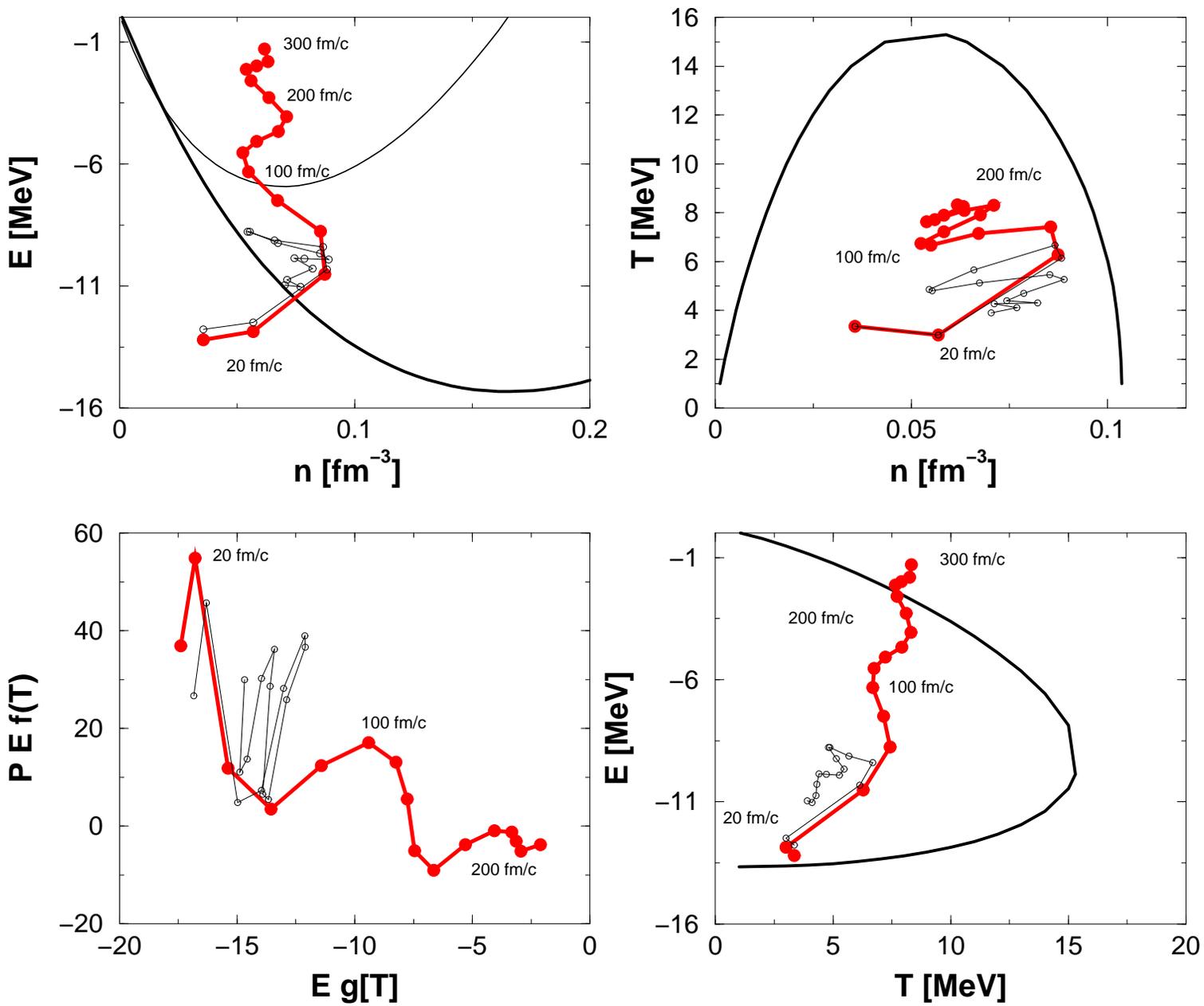,height=20cm,angle=0}}
\caption{The same dynamical trajectories as in figure
  (\protect\ref{xesn25}) but for a reaction $^{56}Ni$ on $^{179}Au$ at $25$MeV
  lab energy. }\label{niau25}
\end{figure}

We might now search for a situation where we have the opposite extreme
that is we search for a reaction with as less as possible surface
emission instability and as much as possible spinodal decomposition. For this
reason we might think on asymmetric reactions since the different
sizes of the colliding nuclei might suppress the surface
emission. Indeed as can be seen in figure \ref{niau25} for an
asymmetric reaction of $Ni$ on $Au$ at $25$MeV lab-energy with nearly the same total charge as in the
reaction before that the surface emission instability is less pronounced while
the spinodal instability is much more important. There appears even a
third minima showing that the matter suffers spinodal decomposition
perhaps more than once if the bombarding energy is low enough and a
long oscillating piece of matter is developing.

\begin{figure}
\centerline{\psfig{file=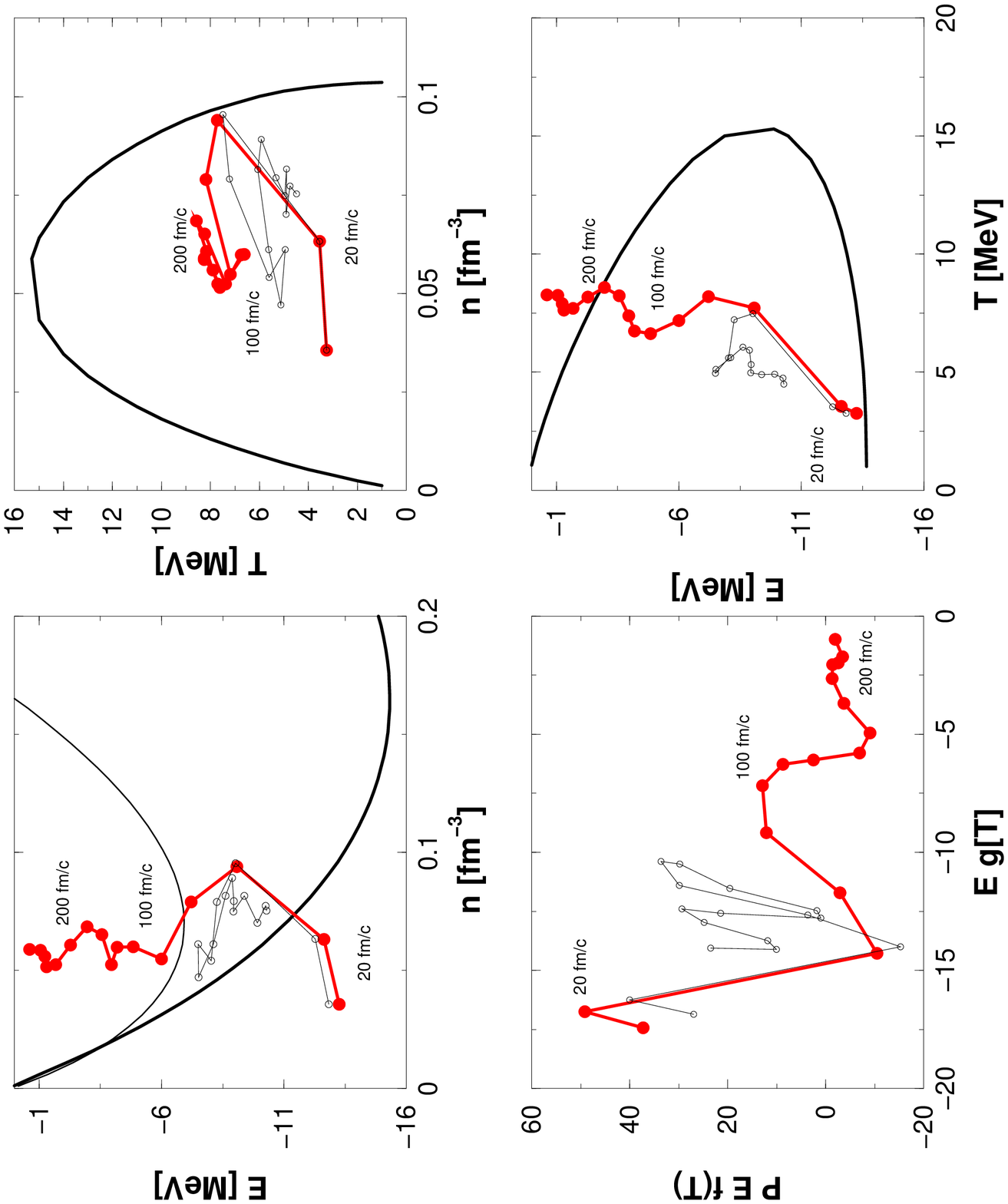,height=20cm,angle=0}}
\caption{The same figure as \protect\ref{niau25} but for $33$MeV lab energy.}\label{niau33}
\end{figure}

\begin{figure}
\centerline{\psfig{file=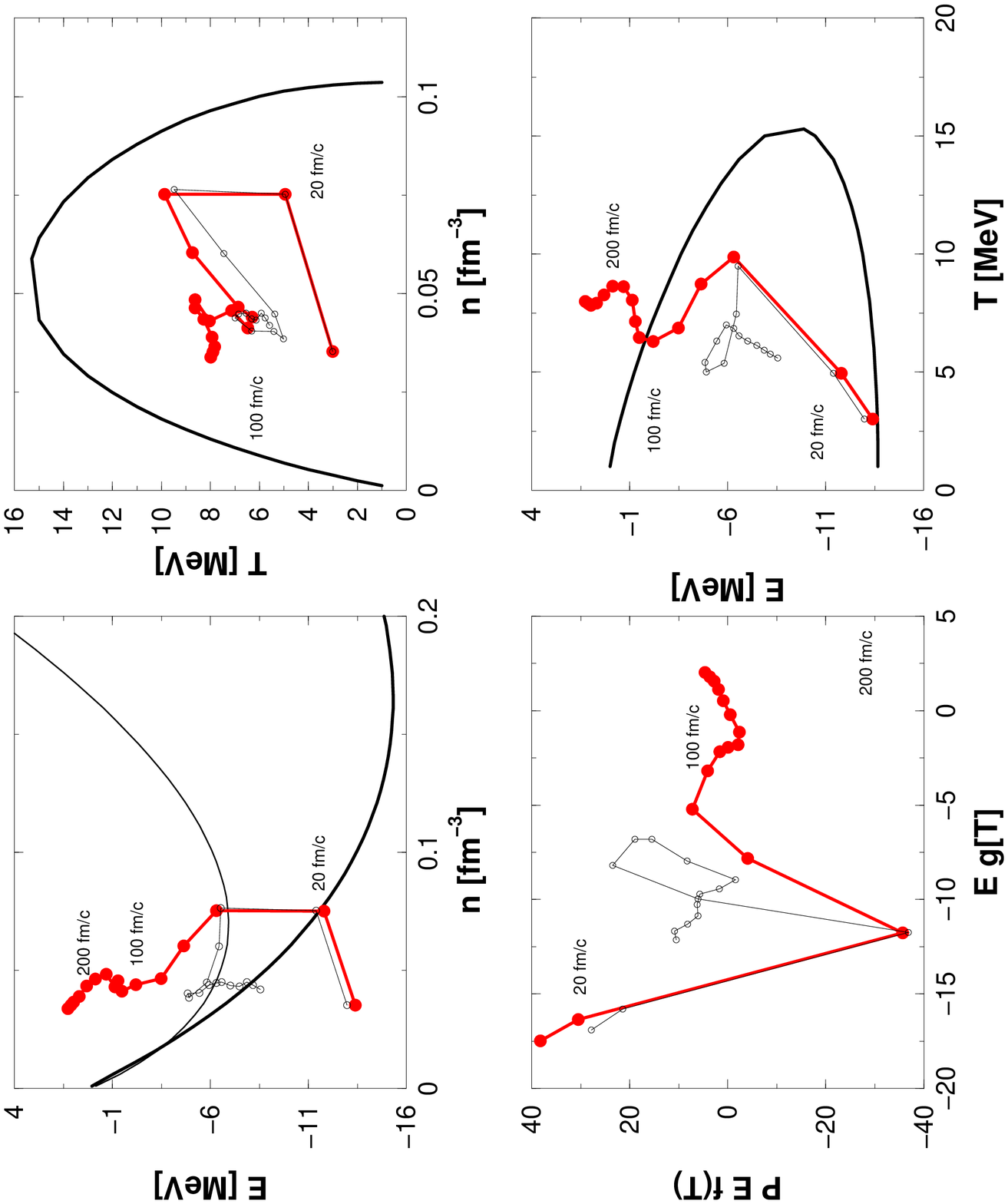,height=20cm,angle=0}}
\caption{The same figure as \protect\ref{niau25} but for $50$MeV lab
  energy. Please note that the time point of highest compression is
  between $20$fm/c and $40$fm/c and not resolved.}\label{niau50}
\end{figure}

The higher bombarding energies now show the same qualitative effect in
that it pronounces the surface emission instability and reduces the importance
of the spinodal decomposition as can be seen in figures \ref{niau33}
and \ref{niau50}. Please note that much smaller compression densities and
temperatures are reached in these reactions compared to the more symmetric
case of $Xe$ on $Sn$.

\section{Summary}

The nonlocal kinetic theory leads to a different nonequilibrium
thermodynamics compared to the local BUU. We see basically a higher
energetic particle spectra and a higher temperature of $2$MeV. This is
attributed to the conversion of two-particle correlation energy into
kinetic energy which is of course absent in local BUU scenario.

By constructing a temperature independent combination of
thermodynamical variables we are able to investigate the signals of
phase transitions under iso-nothing conditions. Two mechanisms of instability have been identified:
surface emission instability and spinodal decomposition. We predict for the
currently investigated reactions seen in table \ref{tt}
which effect should be the leading one for matter decomposition.
\parbox[]{10cm}{
\begin{table}
\begin{tabular}[t]{|l ||c|c|c|}
&25 MeV&33 MeV&50 MeV\\
\hline
&&&\\
$^{58}_{28}$Ni$$ + $^{197}_{79}$Au$$  & {S}   &{C}  {
  S}  &{ C} ({ S}) \\&&&\\
$^{129}_{54}$Xe$$ + $^{119}_{50}$Sn$$ & { C} { S} &{ C} ({ S})
&{ C}     \\&&&\\
\end{tabular}
\begin{tabular}[t]{|l ||c|c|c|}
&15 MeV&33 MeV&60 MeV\\
\hline
&&&\\
$^{157}_{64}$Gd$$ + $^{238}_{92}$U$$  &--  &{ C}  { S}  &{ C}  \\&&&\\
$^{181}_{73}$Ta$$ + $^{197}_{79}$Au$$ &{ C} { S} &{ C} ({ S}) &{
  C}  \\&&&\\
\end{tabular}
\vspace{2ex}
 \caption{The prediction of the leading mechanisms of matter
   disintegration for two reactions with equal total charge but
   asymmetric entrance channels. Surface compression is denoted by $C$
   and spinodal decomposition by $S$.}\label{tt}
\end{table}
}
In the reactions with bombarding energies higher than the Fermi energy the 
fast surface eruption happens outside the spinodal region. 
For even higher energies there is not enough time for the system to rest at
the spinodal region. The trajectories simply move through the spinodal and
the 
system decays before it comes to an equilibrium - like state inside
the spinodal region.

\acknowledgements
B. Tamain is thanked for reading the manuscript and helpful
comments. Especially I am obliged to S. Toneev who brought the idea
of softest point \cite{HS95} to my attention.
Also the discussions with R. Bougault, F. Gulminelli, M. P{\l o}szajczak
and J. P Wieleczko are gratefully acknowledged.

%\bibliography{kmsr,kmsr1,kmsr2,kmsr3,kmsr4,kmsr5,kmsr6,kmsr7,delay2,delay3,spin,refer}
%\bibliographystyle{prsty}

\newpage
\onecolumn

\begin{figure}
\centerline{\psfig{file=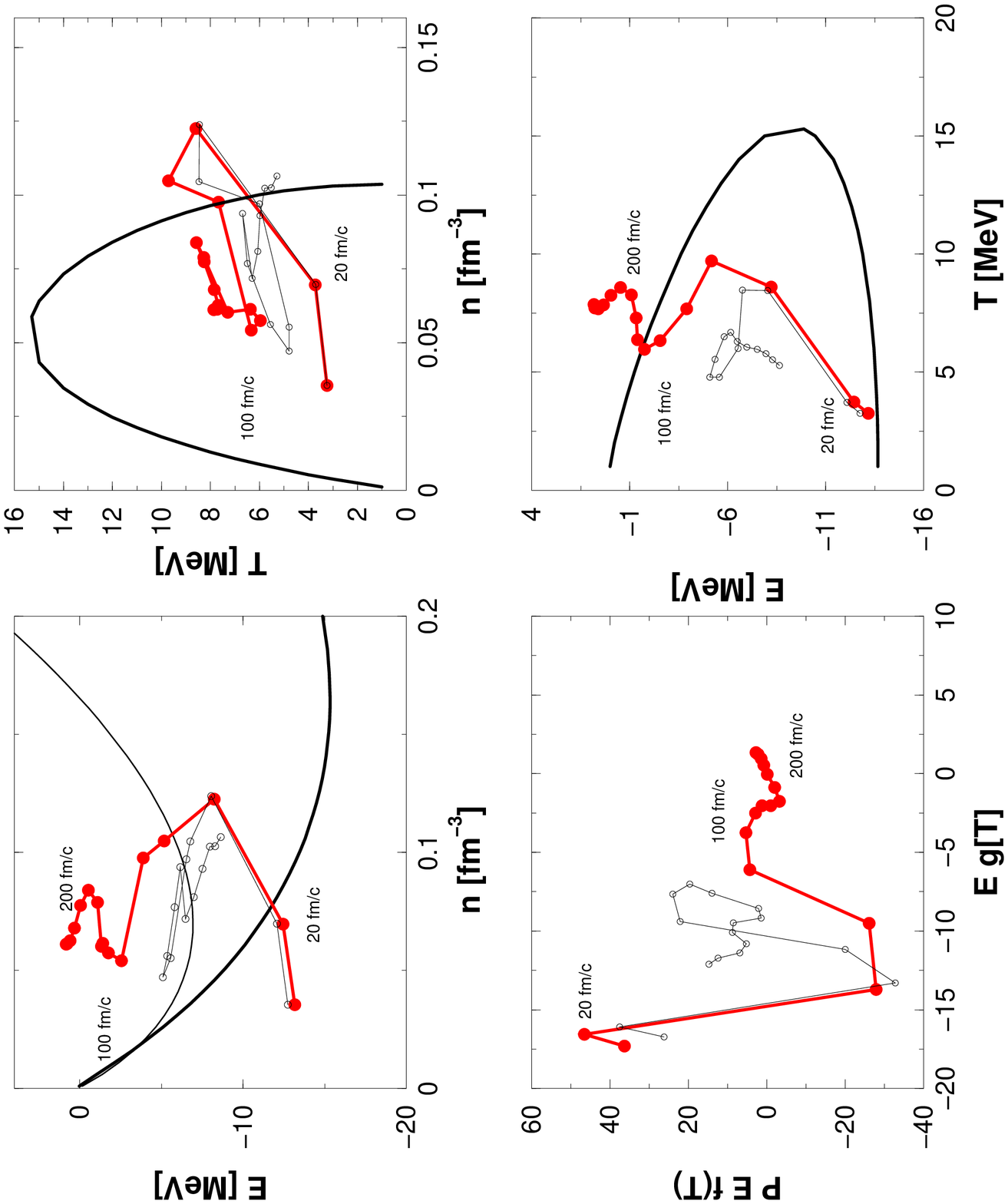,height=20cm,angle=0}}
\caption{The same figure as \protect\ref{xesn25} but for $33$MeV lab energy.}\label{xesn33}
\end{figure}

\end{document}